\documentclass{pazh}

\usepackage{graphicx}
\usepackage{latexsym}

\def\farcs{\hbox{$.\!\!^{\prime\prime}$}}
\def\fdeg{\hbox{$.\!\!^{\circ}$}}

\begin{document}

\titlerunning{Synchrotron superbubble}

\authorrunning{Lozinskaya et~al.}

\title{Synchrotron Superbubble in the Galaxy IC~10: The Ionized
Gas Structure, Kinematics, and Emission Spectrum}

\author{T.A. Lozinskaya$^{1}$ \and A.V. Moiseev$^{2}$ \and N.Yu. Podorvanyuk$^{1}$ \and A.N. Burenkov$^{2}$}

\institute{Sternberg Astronomical Institute, Universitetskii pr.
13, Moscow, 119992 Russia \and Special Astrophysical Observatory
RAS, Nizhnii Arkhyz, Karachai-Cherkessian Republic, 357147 Russia}

\offprints{T.A.  Lozinsksaya, \email{lozinsk@sai.msu.ru}}

\abstract{
We have investigated the structure, kinematics, and emission spectrum of the
ionized gas in the synchrotron superbubble in the irregular galaxy IC~10 based
on observations with the 6-m Special Astrophysical Observatory telescope with
the SCORPIO focal reducer in three modes: direct imaging in the
[SII]$\lambda(6717+6731)$~\AA\ lines, long-slit spectroscopy, and spectroscopy
with a scanning Fabry--Perot interferometer. We have identified a bright (in
the [SII] lines) filamentary optical shell and determined its expansion
velocity, mass, and kinetic energy. The nature of the object is discussed.
}
\maketitle

\section{Introduction}

The dwarf irregular galaxy IC~10 is widely used to investigate the structure,
kinematics, and emission spectrum of the interstellar medium in starburst
regions. Multiple ionized and neutral shells and supershells, arc and ring
structures with sizes from 50 to 800\mbox{--}1000~pc, and a diffuse ionized gas
component are observed in this galaxy (Zucker 2000; Wilcots and Miller 1998;
Gil~de~Paz et~al. 2003; Leroy et~al. 2006; Chyzy et~al. 2003; and references
therein). Its stellar population suggests a recent starburst
($t=4$--$10$~Myr) and an older starburst ($t>350$~Myr) (Hunter 2001;
Zucker 2002; Massey et~al. 2007).

Two features distinguish IC~10 among other dwarf starburst galaxies and make
its studies particularly interesting:

(1)~The anomalously high number of Wolf--Rayet (WR) stars, which is a factor
of~20 greater than that in the Large Magellanic Cloud. The space density of
WR~stars reaches~11 per square kiloparsec (Massey et~al. 1992; Richer et~al.
2001; Massey and Holmes 2002; Crowther et~al. 2003; Massey et~al. 2007; Vacca
et~al. 2007; and references therein). This is the highest density of WR~stars
among the dwarf galaxies comparable to that in massive spiral galaxies. For a
normal initial mass function, such a high density of WR~stars is indicative of
an almost ``simultaneous'' current starburst affecting the bulk of the galaxy.

(2)~The so-called synchrotron superbubble discovered by Yang and
Skillman~(1993) and, as it seems to us, has not yet been explained
exhaustively. The authors associated this extended source of
nonthermal radio emission with the explosions of about ten
supernovae. Given that the characteristic detectability time of
the synchrotron radio emission from a supernova remnant is
$\sim10^{5}$~yr, these explosions should also have occurred almost
simultaneously. The model of multiple supernova explosions is also
used in present-day papers (see, e.g., Bullejos and Rosado 2002;
Rosado et~al. 2002; Thurow and Wilcots 2005). Recently Lozinskaya
and Moiseev (2007) offered an alternative explanation for the
nature of the synchrotron superbubble: a hypernova explosion.

The synchrotron nature of the radio emission from the superbubble is confirmed
by the high degree of its polarization revealed by Chyzy et~al.~(2003).
Figure~2 from the above paper shows that the polarization vector is parallel to
the dense dust layer observed here.

Note that, in fact, the observations by Yang and Skillman~(1993) reveal no
shell structure of this radio source, so the term ``superbubble'' was
introduced by the authors based on the nature of the object suggested by them.

In this paper, we investigate in detail the structure, kinematics, and emission
spectrum of the ionized gas in the region of the nonthermal radio source based
on observations with the 6-m Special Astrophysical Observatory (SAO) telescope
with the SCORPIO focal reducer in three modes (direct [SII]~line imaging,
long-slit spectroscopy, and spectroscopy with a scanning Fabry--Perot
interferometer). The data obtained revealed a bright (in [SII] lines)
filamentary shell that could be identified with the synchrotron superbubble. We
determined the expansion velocity, mass, and kinetic energy of the optical
shell and discuss the nature of the object.

In the succeeding sections, we describe the observing and data reduction
techniques, present and discuss the results obtained, and summarize our main
conclusions.

The published estimates of the color indices and distance for IC~10 vary
greatly due to the low Galactic latitude of the object,
$b=$3\fdeg3 (Demers et~al. 2004). In our paper, we use the values
of $E(B-V)=0.95^{m}$ and $(m-M)_{\textrm{0}}=24.48\pm
0.08^m$ refined by Vacca et~al.~(2007), corresponding to a distance
of $r=790$~kpc to IC~10 and an angular scale of $\simeq 4$~pc/$''$.

All of the radial velocities in this paper are heliocentric.

\begin{table*}
\caption{Log of photometric observations}
\begin{tabular}{l|c|c|c|c|c}
\hline \multicolumn{1}{c|}{Range} & Date &$\lambda_c$, \AA & FWHM, \AA &
$T_{\textrm{exp}}$, s &
Average seeing, $''$ \\
\hline
$\textrm{[SII]}$          & Feb.~14/15, 2007& 6740     &\phantom{0}50 &5600          & 1.1--1.5 \\
$\textrm{[SII]}$~continuum& Feb.~14/15, 2007& 7040     &  210         &1650          & 1.1--1.5 \\
\hline
\end{tabular}
\end{table*}

\begin{figure*}
\includegraphics[scale=.8]{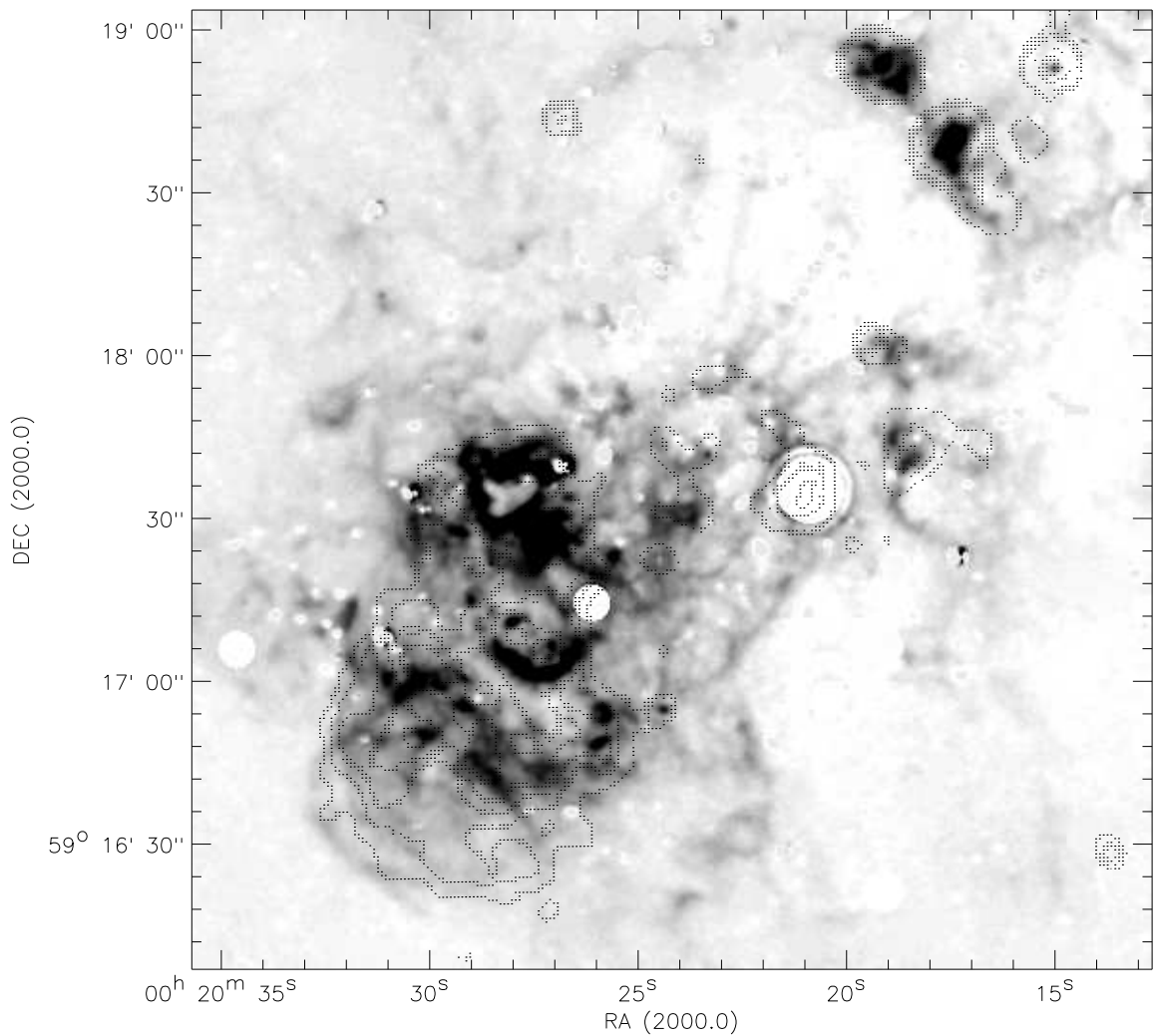}
\caption{[SII]$\lambda(6717+6731)$~\AA\ lines image of the
southeastern part of the galaxy obtained with the 6-m telescope
with 20-cm radio emission contours from Yang and Skillman~(1993)
superimposed.}
\end{figure*}

\section{Observations and data reduction}

\subsection{Interferometric Observations}

Interferometric observations of IC~10 were performed on September 8/9, 2005, at
the prime focus of the 6-m telescope using the SCORPIO focal reducer; the
equivalent focal ratio of the system was $F/2.6$. SCORPIO was described by
Afanasiev and Moiseev~(2005) and on the Internet
(\verb"http://www.sao.ru/hq/lsfvo/devices.html"); the SCOPRIO capabilities in
interferometric observations were also described by Moiseev~(2002).

We used a scanning Fabry--Perot interferometer (FPI) operating in the 501st
order at the H$\alpha$ wavelength. The spacing between the neighboring orders
of interference, $\Delta\lambda=13$~\AA, corresponded to a region free from
order overlapping of $\sim600$~km~s$^{-1}$ on the radial velocity scale. The
FWHM of the instrumental profile was $\sim0.8$~\AA, or $\sim35$~km~s$^{-1}$.
Premonochromatization was performed using an interference filter with
$\textrm{FWHM} =13$~\AA\ centered on the H$\alpha$ line. The detector was an
EEV~42\mbox{--}40 $2048\times2048$-pixel CCD~array. The observations were
performed with $2\times2$-pixel binning to reduce the readout time. In each
spectral channel, we obtained $1024\times1024$-pixel images at a scale of
$0.35''$ per~pixel; the total field of view was $6.1'$.

During our observations, we successively took 36~interferograms of
the object for various FPI plate spacings, so the width of the
spectral channel corresponded to $\delta \lambda=0.37$~\AA, or
17~km~s$^{-1}$ near~H$\alpha$. To properly subtract parasitic
ghosts from the galaxy's numerous emission regions, the
observations were performed for two different orientations of the
instrument's field of view. The total exposure time was 10\,800~s;
the seeing (the FWHM of field-star images) varied within the range
0\farcs8--1\farcs3.

We reduced the observations using software running in the IDL environment
(Moiseev~2002). After the primary reduction, the observational data were
represented as $1024\times1024\times36$ data cubes; here, a 36-channel spectrum
corresponds to each pixel. The final angular resolution (after smoothing during
the data reduction) was $\sim$1\farcs2. The formal accuracy of the
constructed wavelength scale was about 3--5~km~s$^{-1}$.

\subsection{Images in Emission Lines}

Images in the [SII]~$\lambda6717,6731$~\AA\ doublet emission lines were
obtained on the 6-m telescope with the SCORPIO focal reducer. A log of
photometric observations is given in Table~1, which lists the dates of
observations, the central wavelengths~($\lambda_{\textrm{c}}$) and FWHMs of the
filter used, the total exposure times $(T_{\textrm{exp}})$, and the average
seeing. This table includes a filter whose passband contained the
[SII]~$\lambda6717,6731$~\AA\ emission lines and a filter centered on the
continuum near these lines.

The detector was an EEV42\mbox{-}40 CCD that provided a scale of $0.36''$
per~pixel (in $2\times2$ readout mode). The data were reduced using a standard
procedure. When constructing the maps of ``pure'' [SII] line emission, we
subtracted the continuum image from the images in the filter containing the
emission lines and continuum.

A monochromatic image in H$\alpha$ was constructed by integrating the flux in
this emission line in the FPI spectra (see the previous section).

\subsection{Long-Slit Spectroscopy}

Spectroscopic observations of IC~10 were performed with the same
SCORPIO instrument operating in the mode of a long-slit
spectrograph with a slit about $6'$ in length and $1''$ in width.
The scale along the slit was $0.36''$ per~pixel. The total
exposure time was 3600~s at a seeing of about 1\farcs4.
We used the volume phase holographic grating VPHG1200R, which
provided the spectral range $\Delta\lambda=5650$--$7350$~\AA\
that included the H$\alpha$, [NII]~$\lambda 6548,6583$~\AA, and
[SII]~$\lambda6717,6731$~\AA\ emission lines. The spectral
resolution estimated from the widths of night-sky lines was about
6~\AA. The data were reduced in a standard way; the
spectrophotometric standard $\textrm{BD}+25d4655$ observed
immediately after the object almost at the same zenith distance
was used for energy calibration. The intensities and radial
velocities of the emission lines were determined by means of
single-component Gaussian fitting. The accuracy of the absolute
velocity measurements estimated from night-sky lines was
$\sim10$--20~km~s$^{-1}$.

\begin{figure*}
{\large \textbf{a)}}\\
\includegraphics[scale=.7]{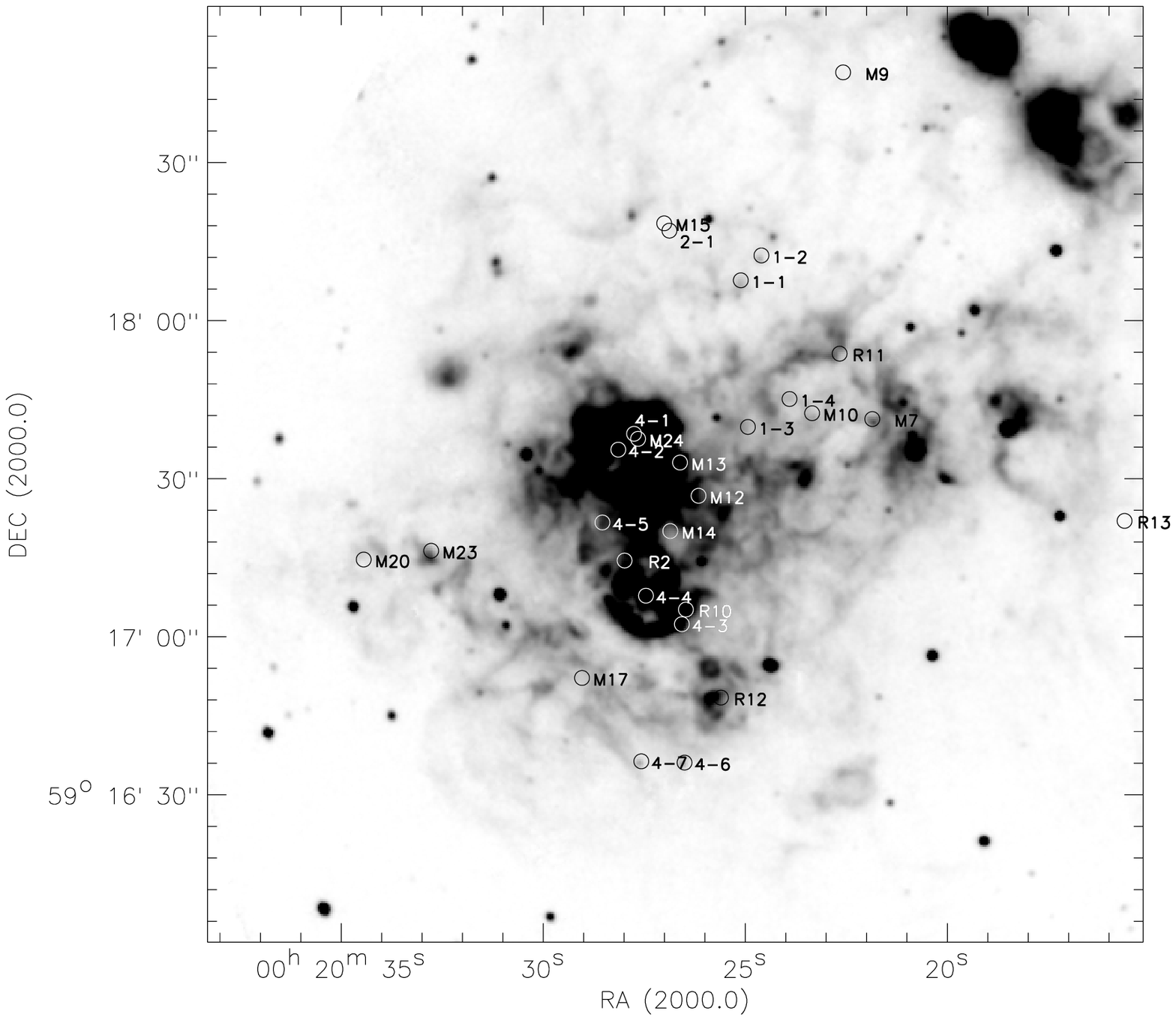}\\
{\large \textbf{a)}}\\
\includegraphics[scale=.7]{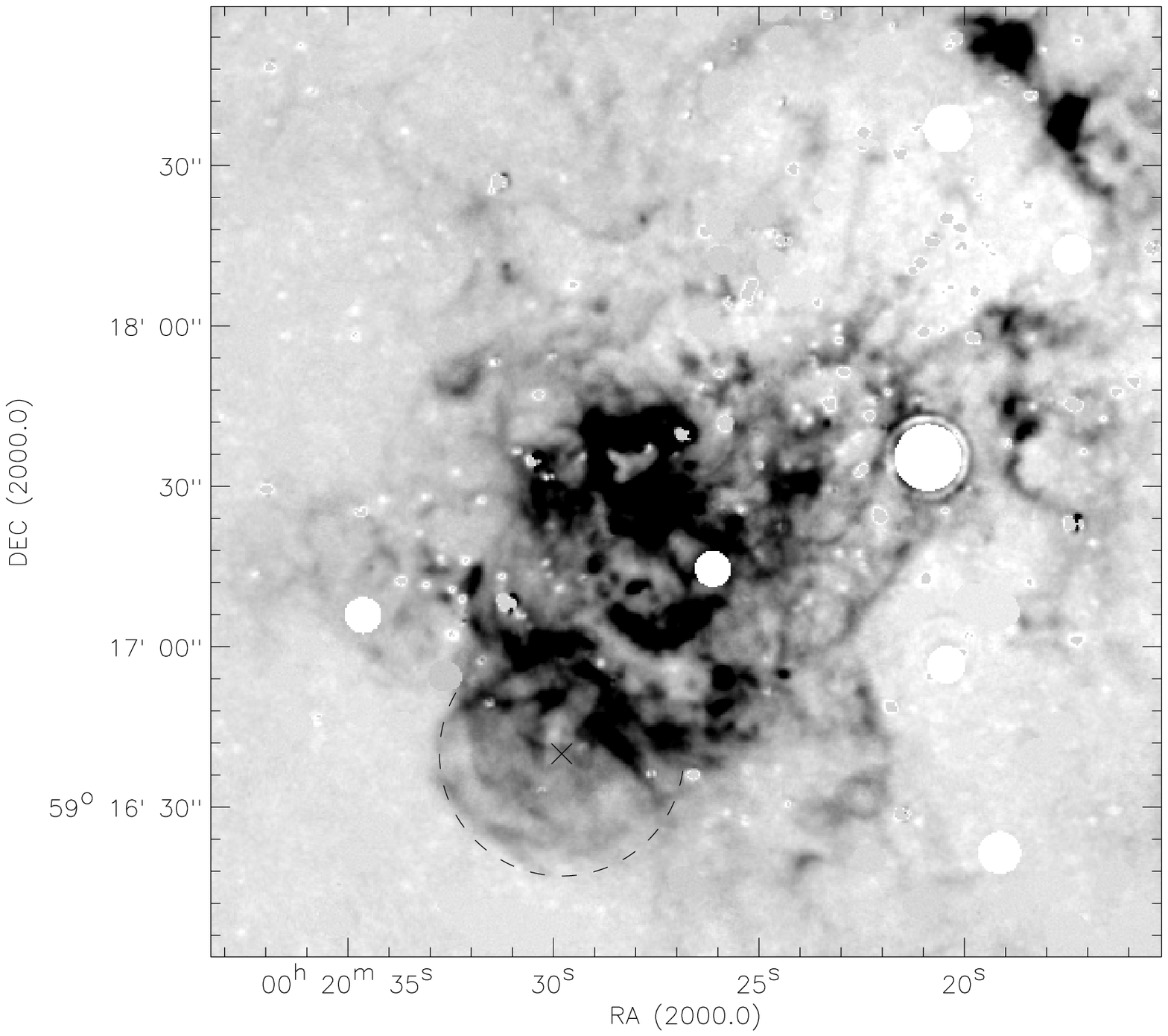}
\caption{Images of the synchrotron superbubble region \textit{(a)} in H$\alpha$ based on
our FPI observations and \textit{(b)} in the [SII]~$\lambda(6717+6731)$~\AA\ lines (the
continuum was subtracted). The positions of spectroscopically confirmed
WR~stars from the lists by Royer et~al. (2001) (denoted by the letter~R) and
Massey and Holmes (2002) (denoted by the letter~M) as well as star clusters
from the list by Hunter~(2001) (denoted by two numerals) are shown.}
\end{figure*}

\begin{figure*}
{\large \textbf{a)}}\\
\includegraphics[scale=.7]{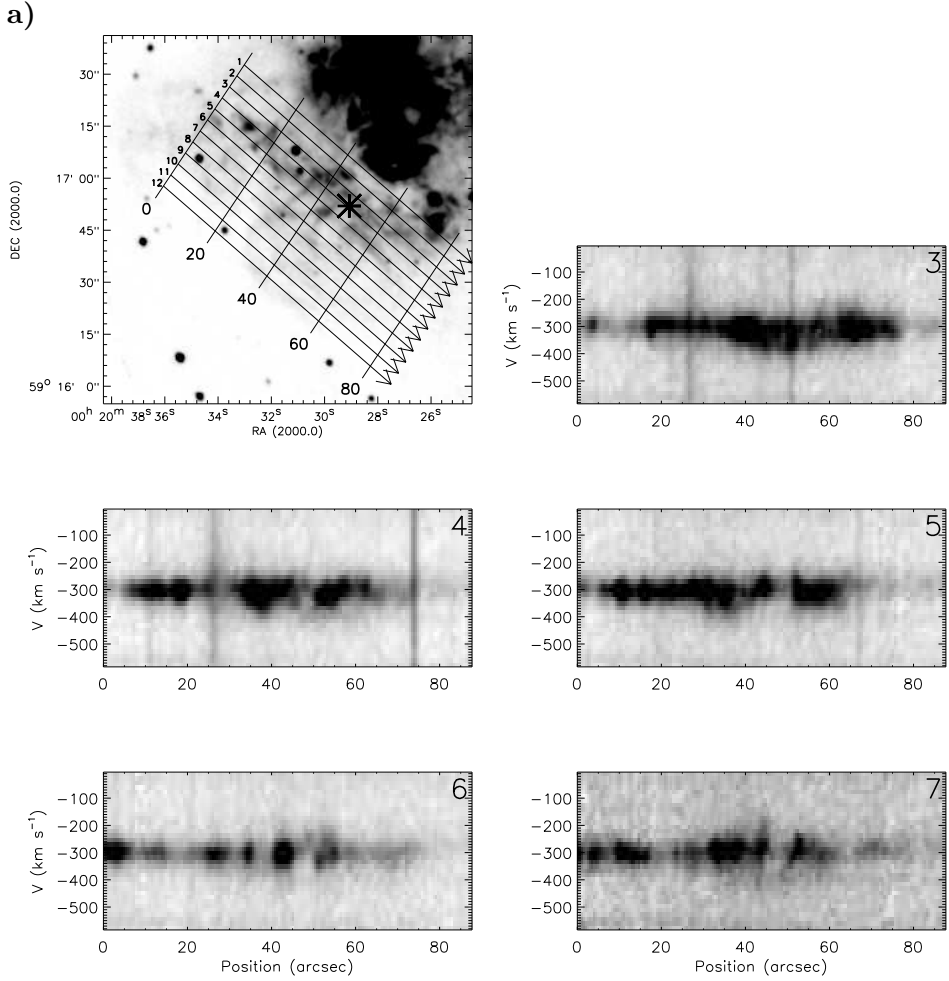}\\ {\large \textbf{a)}}\\
\includegraphics[scale=.7]{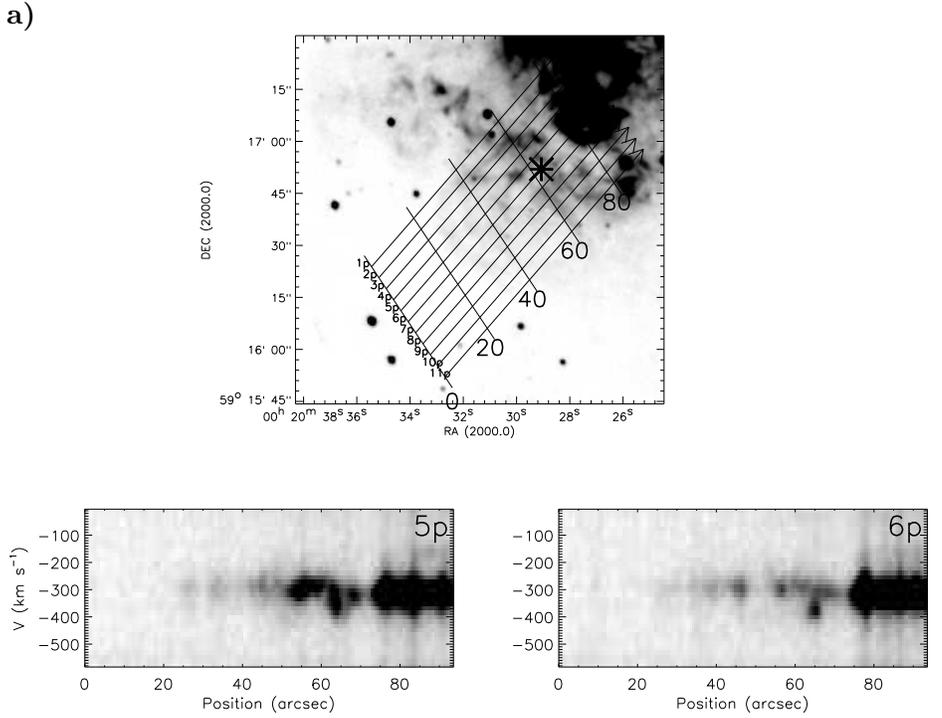}
\caption{Localization of the two systems of scans for which the
$P$/$V$~diagrams were constructed from our FPI observations in
H$\alpha$ and some of the $P$/$V$~diagrams. The scans are marked
up in arsec: \textit{(a)} 12~scans in directions parallel to the dust
layer; \textit{(b)} 11~scans in a perpendicular direction. The asterisk marks the position of the star~WR~M17. }
\end{figure*}

\begin{figure*}
\includegraphics[scale=0.7]{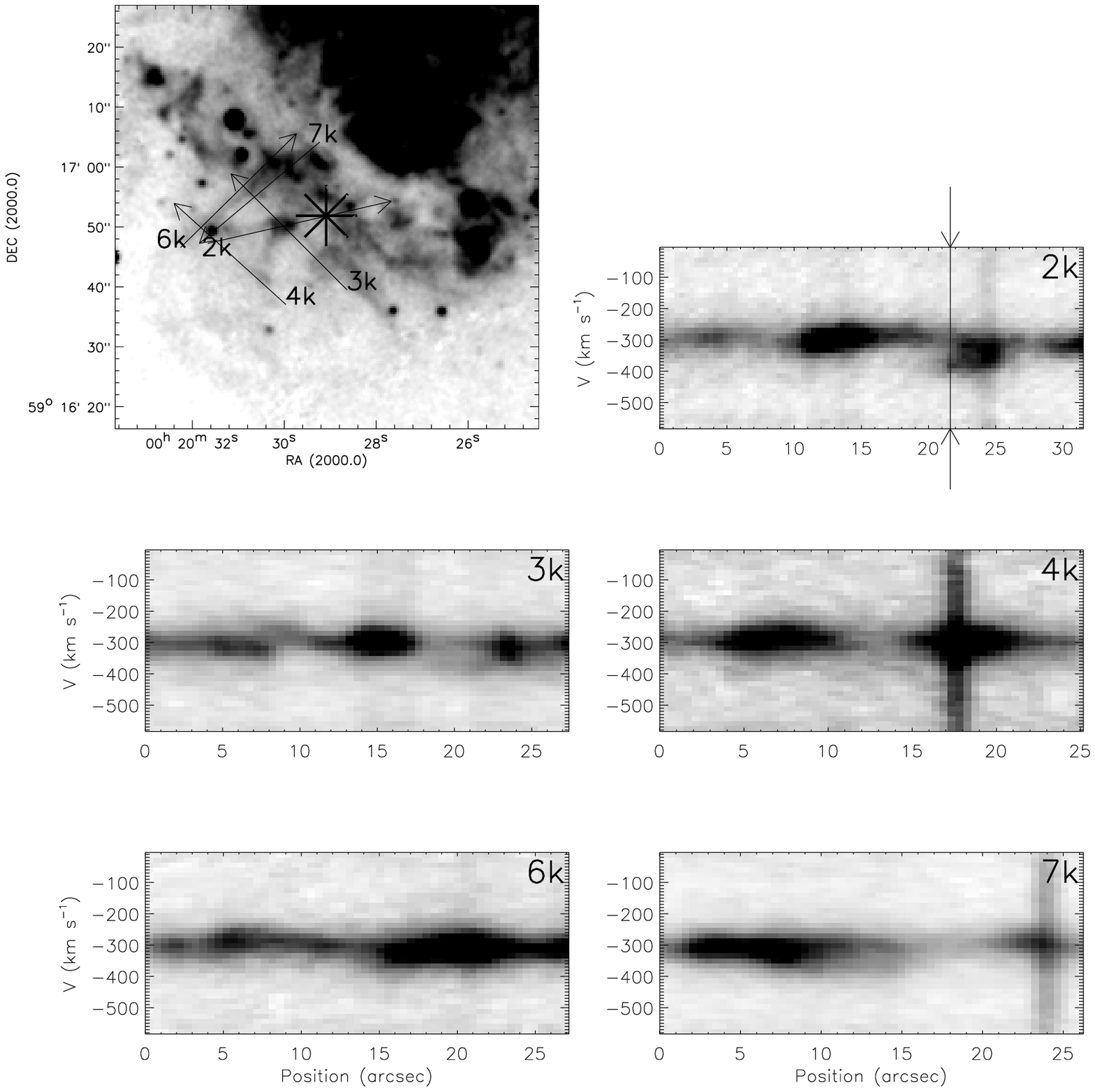}
\caption{Localization of several scans and the corresponding
$P$/$V$~diagrams in the vicinity of the X-ray source~X\mbox{-}1, a
black hole in a pair with the WR star~M17.  The arrows on the
diagram 2k indicate the  star WR M17 location.}
\end{figure*}

\section{Results of observations}

\subsection{Morphology of the Region in the H$\alpha$ and [SII] Lines}

Figure~1 shows our [SII]~$\lambda(6717+6731)$~\AA\ line image of the
southeastern part of IC~10 with 20-cm radio emission contours from Yang and
Skillman~(1993) superimposed. The synchrotron superbubble is located southeast
of the dust lane that is clearly seen in the galaxy's optical images.

Figure~2 compares the H$\alpha$ and [SII] line images of the region. Figure~2a
shows the synchrotron superbubble region in H$\alpha$ based on our FPI
observations. Also shown here are the positions of spectroscopically confirmed
WR stars from the lists by Royer et~al.~(2001) and Massey and Holmes~(2002) as
well as star clusters in the region from the list by Hunter~(2001). Figure~2b
presents our image of the region in the sulfur emission lines. The [SII]
doublet lines are known to be optimal for revealing the emission from the gas
behind the front of a shock propagating with a velocity characteristic of old
supernova remnants. Indeed, a filamentary shell about $40''$--$ 44''$ in
size with the center at $\alpha_{(2000)} =
0^{\textrm{h}}20^{\textrm{m}}29^{\textrm{s}}$, $\delta_{(2000)}
=59^{\circ}16'40''$ is seen most clearly in the [SII] line image of the galaxy.

The shell is morphologically inhomogeneous in the plane of the sky: a brighter
diffuse symmetric spherical structure about $30''$ in size with the center at
$\alpha_{(2000)} = 0^{\textrm{h}}20^{\textrm{m}}30^{\textrm{s}}$,
$\delta_{(2000)} =59^{\circ}16'44''$ and a filamentary structure about $44''
\times (30$--$40)''$ in size with the center at $\alpha_{(2000)} =
0^{\textrm{h}}20^{\textrm{m}}30^{\textrm{s}}$,
$\delta_{(2000)}=59^{\circ}16'36''$ elongated in the same direction as the dust
layer are noticeable in the [SII] lines. These can be a dumbbell-type structure
or an aspherical shell with a nonuniform brightness. Both variants can be
naturally explained by a nonuniform density of the ambient interstellar medium,
as evidenced by the presence of a dense HI and CO cloud (see the text below)
and a dust layer. This optical shell structure, which is bright in the
[SII]~$\lambda(6717+6731)$~\AA\ lines, agrees in localization and size with the
the radio shell (see Fig.~1). Therefore, it can be identified with the optical
emission from the synchrotron superbubble.

At least three more fainter filamentary shell structures that go far beyond the
synchrotron superbubble are observed northeast, west, and southwest of this
bright optical shell.

\begin{figure}
\includegraphics[scale=0.5]{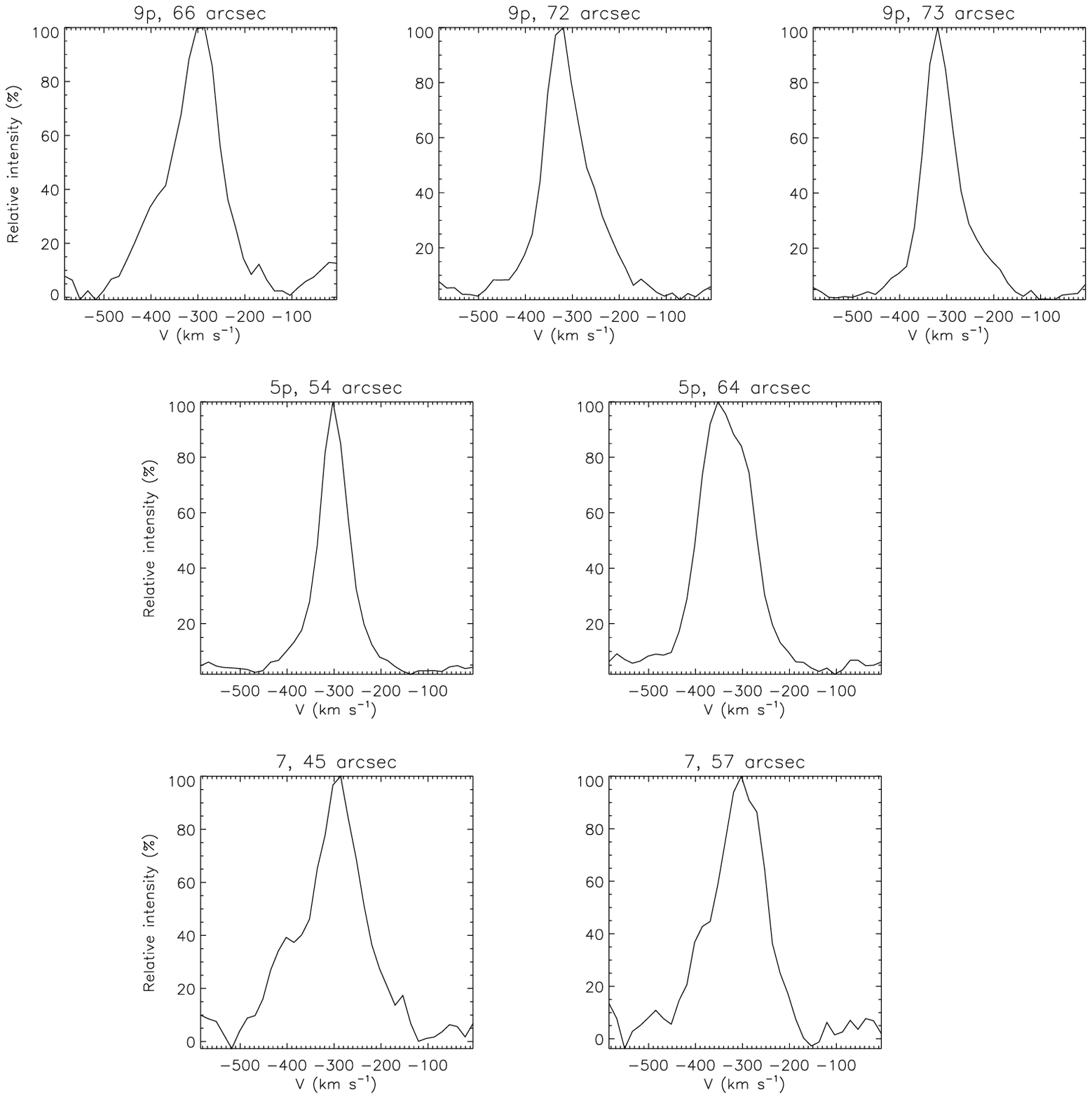}
\caption{H$\alpha$ profiles for scans nos.~9p (in a region of 66,
72, and 73 arcsec),  5p (54 and 64 arsec), and~7 (45 and 57 arcsec) shown in
Fig.~3.}
\end{figure}

\begin{figure*}
\includegraphics[scale=0.8]{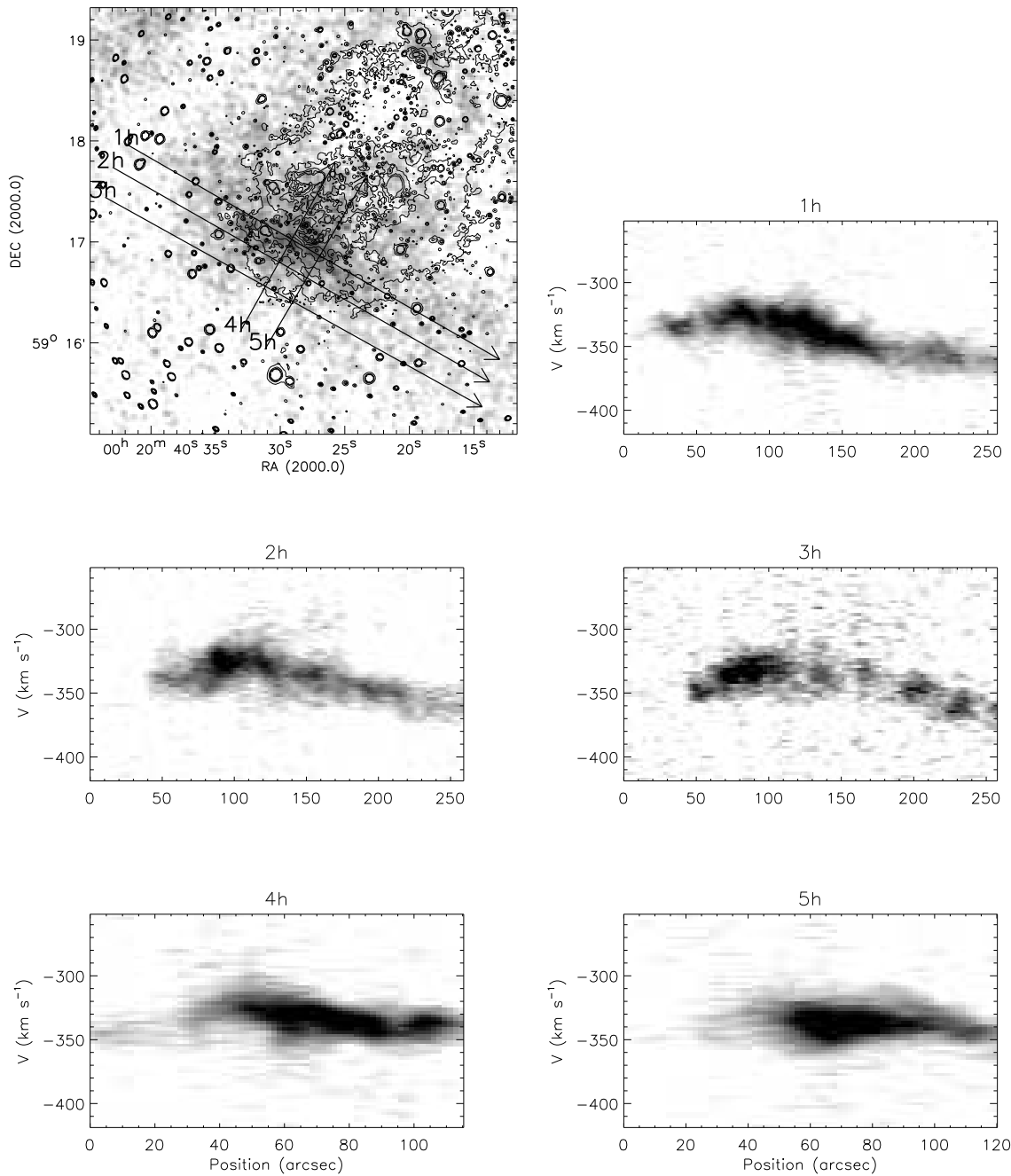}
\caption{Localization of the HI scans passing through the synchrotron
superbubble and the $P$/$V$~diagrams that we constructed from the 21-cm
observations by Wilcots and Miller~(1998). The shades of gray indicate the HI
brightness distribution; the isophotes indicate the
[SII]~$\lambda(6717+6731)$~\AA\ line image.}
\end{figure*}

The [SII] line image of the galaxy from Rosado et~al.~(1999) is also indicative
of an intense emission in the region. We do not associate the arc structure far
beyond the western boundary of the radio source that was mentioned by these
authors with the synchrotron superbubble, because there is no detectable radio
emission here.

As we see, the optical emission is characterized by a filamentary
structure both in the synchrotron superbubble and in its immediate
vicinity. The wind from at least four WR stars can be important in
producing the filamentary structure of the region: M20 and~M23 are
located in the ``northeastern'' shell, M17 is located in the shell
identified with the synchrotron radio source; R12 is located in
the bright part of the ``western'' shell. The star~M17 is
currently believed to be a component of an X-ray binary: a WR star
in a pair with a black hole --- the brightest X-ray source~X-1 in
the Galaxy (Brandt et~al. 1997; Bauer and Brandt 2004; Wang et~al.
2005). As we showed in (Lozinskaya and Moiseev~2007), this object
can represent the compact remnant of a hypernova whose explosion
gave rise to the synchrotron superbubble.

The contribution of two star clusters, 4\mbox{-}6 and 4\mbox{-}7, from the list
by Hunter~(2001) can also be important in producing multiple shells in this
region of the galaxy.

\subsection{Ionized Gas Kinematics in the Synchrotron Superbubble}

Based on our H$\alpha$ observations with a scanning FPI, we investigated the
kinematics of the ionized hydrogen in the entire region southeast of the dust
layer. Taking into account the complex multishell filamentary structure of the
region, the presence of six stellar wind sources mentioned above, and the
possible presence of several supernova explosions, we investigated this region
in greatest detail. Diagrams of the gas radial velocity distribution (the
so-called position--velocity or $P$/$V$~diagrams) were constructed along
45~scans crossing the synchrotron superbubble in various directions.

Figures~3a and~3b show the localization of 12~scans in directions parallel to
the dust layer and 11~scans in a perpendicular direction. For the convenience
of orientation, the scans in each figure are marked up in arcsec. Also shown
here are some of the $P$/$V$~diagrams constructed for these scans that allow
high-velocity motions to be revealed in the region.

The mean gas velocity in the galactic region under consideration was estimated
from our FPI H$\alpha$ observations to be about
$-300$--$-310$~km~s$^{-1}$; we revealed a radial velocity gradient from
the northeast to the southwest up to $-320$--$-330$~km~s$^{-1}$. This is
in agreement with the data from Thurow and Wilcots~(2005) ($-315 $--$
-320$~km~s$^{-1}$, as inferred from H$\alpha$ observations) and the data from
Wilcots and Miller~(1998) ($-326$~km~s$^{-1}$, as inferred from 21-cm HI
observations).

Our $P$/$V$~diagrams do not reveal the classical velocity ellipse (the radial
velocities of the near and far sides of an expanding shell decrease in absolute
value along the radius in the plane of the sky). This can be naturally
explained by the inhomogeneity of the synchrotron superbubble, which is
actually represented by knots and filaments both in H$\alpha$ and in the [SII]
lines (see Fig.~2).

At the same time, we see distinct deviations of the velocities of ionized gas
clouds and filaments from the derived mean gas velocity in the shell. Scans
nos.~3 (positions 45--65~arcsec), 4 (37--57~arcsec), 5 (37--53~arcsec), 6
(45--53~arcsec), and 7 (about 47~arcsec) as well as nos.~5p (61--67~arcsec)
and~6p (61--67~arcsec) presented in Fig.~3 demonstrate the largest velocity
deviations.

As we see, all these scans pass through the central part of the superbubble
near the brightest X-ray source X\mbox{-}1 in the galaxy, a black hole in a
pair with the WR star~M17. Since this object is an obvious source of kinetic
energy that affects the gas structure and kinematics in the synchrotron
superbubble, more than 10 additional scans were made in its immediate vicinity
on a larger scale. The localization and corresponding $P$/$V$~diagrams of
several of them are shown in Fig.~4.

The gas kinematics near all WR stars in the galaxy will be investigated in
detail in our next paper (Lozinskaya et~al. 2008). Therefore, we do not
consider here the three more WR stars mentioned above that are located in the
northeast and the southwest outside the synchrotron superbubble.

As follows from Fig.~4, the star~M17 is located in a local
aspherical cavity elongated from the north to the south. The
distance to the outer boundaries of the cavit
y is about~$2''$ north, east, and west of the star and reaches~$5$--$6''$ in
the southeast (8 and 19--23~pc, respectively). The largest
deviations of the mean radial velocity from a characteristic value
of $-300$~km~s$^{-1}$ in the superbubble region are observed in
the bright walls of this cavity: about $-360 $--$
-380$~km~s$^{-1}$ on scans nos.~2k, 4k, and~6k. The mean
velocities of the features with a positive shift relative to the
mentioned characteristic velocity reach $-280
$--$-250$~km~s$^{-1}$ (see scans nos.~2k --- position
14~arcsec, 3k --- 8 and 15~arcsec, 6k
--- 20\mbox{--}22~arcsec, and 7k --- 15~arcsec). These features are in the shape
of bridges in the $P$/$V$~diagrams and correspond to low-brightness regions
inside the cavity.

We emphasize that the velocities mentioned above were determined from the line
peak, i.e., they characterize the brightest knots and filaments. We estimated
the expansion velocity of the superbubble determined by the deviations of the
line peak velocity from its mean value in the bright northern part of the
nebula to be about 50--80~km~s$^{-1}$. Fainter wings at a level of about
20--30$\%$ of the peak intensity are observed in the synchrotron superbubble
region in the velocity range from $- 450 $--$ -420$ to $- 200 $--$
-240$~km~s$^{-1}$.

The H$\alpha$ profiles for scans nos.~9p (in a region of 66, 72, and
73~arcsec), 5p (54 and 64~arcsec), and~7 (45 and 57~arcsec) presented in Fig.~5
can serve as an example. As we see from the figure, some of these profiles
exhibit a double-peaked structure.

\subsection{Neutral Gas Kinematics in the Synchrotron Superbubble}

Based on 21-cm line observations, Wilcots and Miller (1998) investigated in
detail the kinematics of the neutral hydrogen in the galaxy. We reanalyzed the
data cube kindly provided by these authors to present the results in a form
comparable to our studies of the ionized gas kinematics. Our results completely
confirmed the data from the above paper. Figure~6 shows the localization of
five scans passing through the synchrotron superbubble for which we constructed
the $P$/$V$~diagrams for the neutral gas. As follows from the figure, the
neutral hydrogen in the region of the synchrotron radio source clearly reveals
kinematic signatures of the shell expansion --- the classical picture of half
the velocity ellipse corresponding to the receding side of the shell. We
emphasize that the shell structure in the neutral hydrogen distribution in the
region of the synchrotron radio source is unseen; we are talking about an
expanding shell only based on the gas kinematics. The mean expansion velocity
of the HI shell is 25--30~km~s$^{-1}$. The mass of the neutral gas drawn into
this expansion reaches $M\simeq 7\times 10^{5} M_{\odot}$, as estimated by
Wilcots and Miller (1998) (the authors point out that this value is determined
with a low accuracy). The corresponding kinetic energy is
$E_{\textrm{kin}}\simeq 5\times 10^{51}$~erg.

The vicinities of the synchrotron superbubble are the dynamically most active
region of the galaxy in the 21-cm line: such a distinct velocity ellipse is
observed nowhere and the possible expansion velocity of the local HI shells in
IC~10 estimated by Wilcots and Miller~(1998) nowhere exceed 20~km~s$^{-1}$.

\subsection{The Gas Emission Spectrum in the Synchrotron Superbubble}

The localization of the slit spectrogram passing through the synchrotron
superbubble is shown in Fig.~7. For the convenience of identification, the
spectrogram in the figure is marked up in arcsec; the region from $90''$
to~$125''$ corresponds to the superbubble.

The spectrum processing results are presented in Fig.~8; when
analyzing the spectrum, we performed an averaging over 10~pixels
(3\farcs6) along the slit to increase the
signal-to-noise ratio for faint emission-line regions. The error
boxes correspond to $3\sigma$.

As follows from Fig.~8, the line intensity ratio $I$([SII])/$I$(H$\alpha$)
increases significantly in the synchrotron superbubble region, thereby
confirming the above comparison of the images in these lines. The value of
$I([\textrm{SII}])/I(\textrm{H}\alpha) = 0.7$--$1.0$ observed in the
superbubble strongly suggests the emission of the gas behind the shock front.
The criterion $I([\textrm{SII}])/I(\textrm{H}\alpha)\geq 0.6$ commonly used in
revealing supernova remnants was exceeded significantly, given the low
metallicity of IC~10, $Z=0.2$--$0.3 Z_{\odot}$ (Skillman et~al. 1989;
Garnett 1990).

The distribution of the [SII] 6717/6731~\AA\ doublet line intensity ratio
presented in Fig.~8c was used to determine the electron density of the gas in
the synchrotron superbubble. In the [SII] line emission zone, we took
$T_{\textrm{e}} = 10000$~K, a temperature typical of low-excitation zones.
Taking into account the large observational error, we determined the density
$n_{\textrm{e}} \simeq 20 $--$ 30$~cm$^{-3}$ only for the brighter
northern part of the superbubble (the region $90''$--$105''$ along the
spectrogram), where the errors are relatively small.

\begin{figure*}
\includegraphics[scale=0.8]{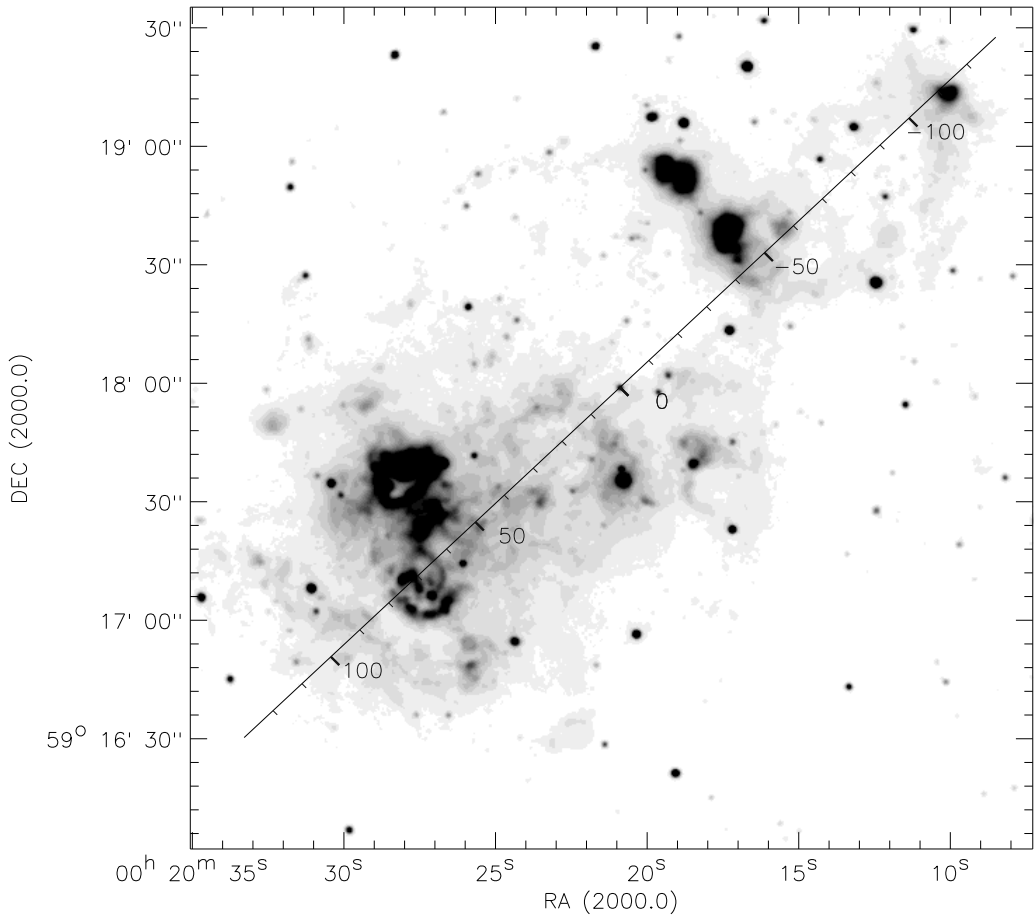}
\caption{Localization of the slit spectrogram passing through the synchrotron
superbubble (image in H$\alpha$+continuum); the spectrogram is marked up in arsec.}
\end{figure*}

\section{Discussion}

Our [SII] line images reveal the optical shell in the region of
the synchrotron radio source much more clearly than the H$\alpha$
images. The total size of the filamentary shell structure, about
$44''$, corresponds to 168~pc. The synchrotron radio source has a
comparable size: about $45$--$ 50''$, i.e., about 170--190~pc
at the distance of 790~kpc assumed here (in Yang and
Skillman~(1993), the size of the radio source, $\geq 250$~pc,
corresponded to a distance of 1.25~Mpc).

The kinematics of the ionized gas in the galaxy was previously
discussed by Bullejos and Rosado~(2002), Rosado et~al.~(2002), and
Thurow and Wilcots~(2005). The first two papers are brief reports
in conference proceedings; an expansion velocity of
50--70~km~s$^{-1}$ is only mentioned in them, which is completely
confirmed by our observations. Thurow and Wilcots~(2005) obtained
90 H$\alpha$ profiles for each of the five galaxy's fields, one of
which covers part of the synchrotron superbubble. The authors
present the radial velocity distribution (we completely confirmed
these measurements) and point out a large line FWHM (about
90~km~s$^{-1}$) in the synchrotron superbubble region.

Our study of the gas kinematics in the synchrotron superbubble region is more
systematic and detailed: we analyzed the position--velocity diagrams for more
than 50 scans of different widths in different directions crossing the entire
superbubble region and its immediate vicinities. As a result of this analysis,
we determined the expansion velocity of the system of bright knots in the
superbubble, $V_{\textrm{exp}} = 50$--$ 80$~km~s$^{-1}$ (the expansion
velocity determined from weak line features is about 100~km~s$^{-1}$).

Based on our long-slit spectrum, we estimated the mean electron
density of the gas in the northern part of the synchrotron
superbubble from the [SII] doublet intensity ratio:
$n_{\textrm{e}} \simeq 20$--$30$~cm$^{-3}$. Thurow and
Wilcots~(2005) found a close value, $n_{\textrm{e}} = 30 $--$
50$~cm$^{-3}$, using the same method, but for a different
orientation of the spectrograph slit.

Taking, as is commonly assumed in supernova remnants, the shell thickness to be
about~0.1 of its radius, we can roughly estimate the mass of the optical shell
and determine its kinetic energy from the expansion velocity found: $M\simeq
4\times 10^{5} M_{\odot}$ and $E_{\textrm{kin}}\simeq (1$--$3)\times
10^{52}$~erg, respectively. Thurow and Wilcots (2005) found a similar kinetic
energy, $E_{\textrm{kin}} = (5$--$6)\times 10^{52}$~erg, from the mean
H$\alpha$ FWHM in the superbubble. A lower kinetic energy of the optical shell,
$E_{\textrm{kin}} = (0.6 $--$ 1.2)\times10^{51}$~erg, is mentioned in
Bullejos and Rosado~(2002) and Rosado et~al.~(2002), probably because these
authors used an underestimated gas density.

As we noted above, the mass of the neutral gas showing signatures of the shell
expansion with a velocity of 25~km~s$^{-1}$, reaches $7\times 10^{5}
M_{\odot}$, as estimated by Wilcots and Miller~(1998). This gives a kinetic
energy $E_{\textrm{kin}}\simeq 4\times 10^{51}$~erg.

No more than $30\%$ of the supernova explosion energy is known (see, e.g.,
Chevalier 1974) to be transferred to the ambient interstellar medium.
Therefore, the total kinetic energy of the ionized and neutral gas found
actually corresponds to the explosions of about ten supernovae, given that the
stellar wind from the host association of these supernovae can provide
approximately the same total kinetic energy input (see the estimates by Yang
and Skillman 1993). Therefore, in present-day works (see, e.g., Bullejos and
Rosado 2002; Rosado et~al. 2002; Thurow and Wilcots 2005), based on the
estimates of kinematic parameters for the ionized gas in the region of the
nonthermal radio source, the authors also adopt the hypothesis of multiple
supernova explosions. Bullejos and Rosado~(2002) and Rosado et~al.~(2002) found
a lower kinetic energy and estimated the necessary number of supernovae to
be~3--6.

In (Lozinskaya and Moiseev~2007) we offered an alternative
explanation for the nature of the synchrotron superbubble: a
hypernova explosion. We pointed out that the binary X-ray source
•\mbox{-}1, which is currently believed (Bauer and Brandt 2004;
Wang et~al. 2005) to be an accreting black hole in a pair with the
WR star~M17, is a possible compact remnant of this hypernova\footnote{
Recently Prestwitch et al. (2007) and Silvereman and Filippenko
(2008)  showed the mass of the black-hole companion to M17
to be 23-35 solar masses and thereby  fully confirmed the
assumption of Lozinskaya and Moiseev (2007) concerning the
Hypernova explosion.}.

\begin{figure}
\includegraphics[scale=0.5]{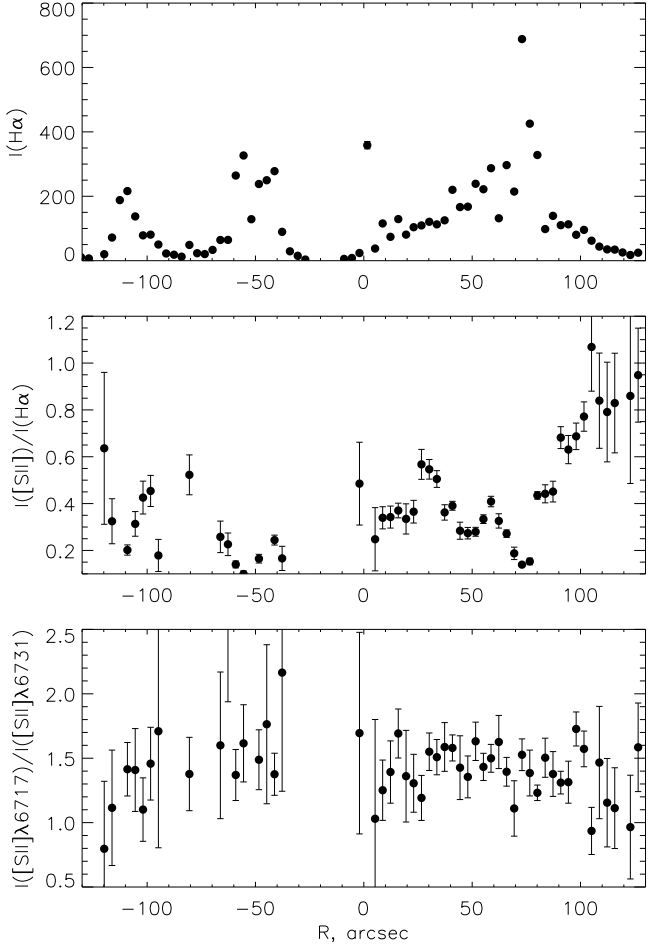}
\caption{ The spectrogram processing results: \textit{top} H$\alpha$
intensity distribution along the slit; \textit{middle} line intensity ratio
$I$([SII]$\lambda(6717+6731)$~\AA)/$I$(H$\alpha$); \textit{bottom} [SII]
doublet line intensity ratio
$I$($\lambda6717$~\AA)/$I$($\lambda6731$~\AA). The region from
$90''$ to $125''$ corresponds to the synchrotron superbubble.
\hfill}
\end{figure}

We emphasize that the kinetic energy of the optical shell found
here can be provided both by a hypernova explosion and by the
model of multiple supernova explosions suggested previously. In
(Lozinskaya and Moiseev 2007), we reached the conclusion that a
hypernova explosion explained better the nature of the synchrotron
superbubble than multiple supernova explosions by analyzing the
radio emission from the object. In this paper, for the first time,
we have taken into account the fact that the explosions of
``recent'' supernovae in the model by Yang and Skillman~(1993)
occur in a tenuous cavity inside the superbubble swept out by the
``first'' supernovae. Our analysis of this effect (Lozinskaya and
Moiseev 2007) showed that subsequent supernova explosions add
little to the radio brightness of the synchrotron superbubble
produced by the first explosions and, accordingly, the necessary
number of explosions in the model of multiple supernovae increases
significantly. Therefore, a hypernova explosion seems a more
plausible formation mechanism of the synchrotron superbubble.

In recent years, interest in the possible hypernova remnants has
increased greatly, because the latter are assumed to be associated
with gamma-ray bursts. However, the kinetic energy of the
superbubble found $E_{0} = (1$--$10)\times10^{52}$~erg can
often be adequately explained not only by a hypernova explosion,
but also in terms of the classical model of the combined action of
stellar winds and multiple supernovae of an OB~association.
Indeed, multiple supernova explosions in the galaxy's local region
suggest the presence of a rich host OB~association. The results of
a detailed analysis of the physical conditions and nature of the
two presumed hypernova remnants in the galaxy M~101 performed by
Lai et~al.~(2001) and Chen et~al.~(2002) can serve as an
illustration. One of them~(MF~83) can be reliably reidentified as
the superbubble swept out and ionized by the stars of four
OB~associations; the second (NGC~5471) can actually be associated
with a hypernova explosion. Note that the object MF~83 is not
associated with the synchrotron radio source, which should have
immediately called into question its identification as being the
result of a hypernova explosion or multiple supernova explosions.
However, this question is not quite unambiguous either: the
presence of a gas ionized by the host association can make the
thermal radio emission in the region predominant.

The anomalously high kinetic energy of some HI shells surrounding classical
supernova remnants (see, e.g., Gosachinskij and Khersonskij~(1985),
Gosachinskij~(2005), and references therein) do not provide unequivocal
evidence for a hypernova explosion either. First, the wind from the supernova
progenitor and/or the stars of the host OB~association is an obvious energy
source of the outer HI shell. Second, the passage from the estimated kinetic
energy of the HI shell to the initial supernova explosion energy is
model-dependent.

Therefore, the main criterion that allows the superbubble in IC~10
investigated here to be reliably considered a remnant of hypernova
explosion is its strong synchrotron radio emission, which, as we
showed in (Lozinskaya and Moiseev 2007), is in better agreement
with a hypernova explosion than with multiple supernova
explosions.

The age of the superbubble determined by its size and the expansion velocity
found, 50\mbox{--}80~km~s$^{-1}$, is $t \simeq (4$--$7)\times10^{5}$~yr,
if the remnant is at the Sedov stage or $t \simeq (3$--$5)\times10^{5}$~yr
for the radiative cooling stage.

This age is a strong argument for the hypothesis of a hypernova explosion. It
takes at least $t \simeq 10^{7}$~yr for the explosions of about ten supernovae
in the galaxy's local region about 200~pc in size.

Note also that the explosions of about ten supernovae in a short time interval
suggest the presence in the region of a rich association of young stars (up to
800~massive stars with $M\geq 10 M_{\odot}$, similar to 30~Dor and~R136, as
estimated by Yang and Skillman (1993)). It is hard to ``hide'' this rich host
association, although it cannot be completely ruled out in principle, given the
strong absorption due to the low Galactic latitude of IC~10 and the location of
the synchrotron superbubble near the densest dust layer in IC~10. Note that
Hunter~(2001) identified two clusters near the southern boundary of the
superbubble, 4\mbox{-}6 and 4\mbox{-}7 in her notation, but their colors were
not determined and there is no age estimate. These are by no means richest
clusters in the galaxy and they are definitely not suitable for the role of the
``hosts'' for the hypothetical multiple supernovae.

In the plane of the sky, the synchrotron superbubble is located in the region
of the densest HI cloud in the galaxy; the densest CO cloud in IC~10 also
coincides with the core of this cloud (see Fig.~4 from Leroy et~al. 2006).

Comparison of the ionized and neutral gas kinematics in the synchrotron
superbubble region reveals a clear ``asymmetry'' in the HII and~HI velocity
distribution. Indeed, the most distinct ``high-velocity'' H$\alpha$~emission
features are shifted toward the negative velocities, while the relatively more
amorphous HI~emission regions represent the receding side of the expanding
shell at the velocities shifted toward the positive values. Based on the
observed ``asymmetric'' kinematics of the ionized and neutral gas, we conclude
that along the line of sight, the synchrotron superbubble is located on the
near side of a dense cloud. The absence of clear morphological signatures of
the HI shell structure in the vicinity of the synchrotron superbubble noted
above most likely stems from the fact that only a small fraction of the HI
column density is involved in this expansion that is clearly revealed by its
kinematics.

The fact that the neutral gas reveals a classical velocity ellipse, while the
ionized gas reveals only separate, rapidly moving knots can be explained by the
homogeneity of the HI region and by the clumpiness of the HII region, which is
clearly seen in the [SII] and H$\alpha$ line images presented above.

\section{Conclusions}

Our [SII] line image reveals the optical shell about 170~pc in size (at a
distance of 790~kpc) that can be identified with the synchrotron superbubble
better than the H$\alpha$ image.

A detailed study of the ionized gas kinematics allowed us to estimate the
characteristic expansion velocity of the bright knots and filaments in the
optical shell: 50--80~km~s$^{-1}$; the expansion velocity determined from weak
line features is $\sim100$~km~s$^{-1}$.

Having estimated the electron density in the northern part of the
optical shell from the [SII]6717/6731~\AA\ ratio, $n_{\textrm{e}}
\simeq 20 $--$ 30$~cm$^{-3}$, by assuming the shell thickness
to be $\sim0.1$ of its radius, we found its mass $M\simeq 4\times
10^{5} M_{\odot}$ and kinetic energy $E_{\textrm{kin}}\simeq
(1$--$3)\times 10^{52}$~erg from the measured expansion
velocity of 50--80~km~s$^{-1}$. This energy is intermediate
between the estimates $E_{\textrm{kin}}\simeq (5$--$6)\times
10^{52}$~erg from Thurow and Wilcots~(2005) and
$E_{\textrm{kin}}\simeq (0.6$--$1.2)\times 10^{51}$~erg from
Bullejos and Rosado~(2002) and Rosado et~al.~(2002). The shell's
energy found corresponds to the explosions of about ten supernovae
plus the stellar wind from their host association, as suggested by
Yang and Skillman~(1993), or a hypernova explosion, as we
suggested in (Lozinskaya and Moiseev~2007).

The shell age, $t \simeq (3$--$7)\times 10^{5}$~yr,
corresponding to the derived expansion velocity is a strong
argument for the hypothesis of a hypernova explosion, since it
takes at least $t \simeq 10^{7}$~yr for the explosions of ten
supernovae in the galaxy's local region.

Comparison of the ionized and neutral gas kinematics leads us to conclude that
the synchrotron superbubble is located on the near side of a dense HI and~CO
cloud observed here.

\begin{acknowledgements}
This work was supported by the Russian Foundation for Basic Research (project
nos.~05-02-16454 and 07-02-00227). The work is based on the observational data
obtained with the 6-m SAO RAS telescope funded by the Ministry of Science of
Russia (registration no. 01-43). We thank A.~Valeev for his help in the
observations and E.~Wilcots, who provided the data cube of HI~observations.
When working on the paper, we used the NASA/IPAC Extragalactic Database (NED)
operated by the Jet Propulsion Laboratory of the California Institute of
Technology under contract with the National Aeronautics and Space
Administration (USA).
\end{acknowledgements}

\textit{Translated by V.~Astakhov}

\end{document}